\documentstyle{article}\begin{document}
\title{Relativistic diffusive motion  in random electromagnetic  fields}
\author{ Z. Haba\\
Institute of Theoretical Physics, University of Wroclaw,\\ 50-204
Wroclaw, Plac Maxa Borna 9, Poland\\email:zhab@ift.uni.wroc.pl}
\date{\today}\maketitle
\begin{abstract} We show that the
relativistic dynamics in a Gaussian random electromagnetic field
can be approximated by the relativistic diffusion of Schay and
Dudley. Lorentz invariant dynamics in the proper time leads to the
diffusion in the proper time. The dynamics in the laboratory time
gives the diffusive transport equation corresponding to the
J\"uttner equilibrium at the inverse temperature
$\beta^{-1}=mc^{2}$.The diffusion constant is expressed by the
field strength correlation function (Kubo's formula).
\end{abstract}
 \section{Introduction}
The attempts to define a relativistic version of diffusion started
a long time ago (see the reviews \cite{deb}\cite{hang}). In
\cite{lopuch}\cite{hakim} it has been shown that relativistic
Markovian diffusion cannot be defined in the configuration space.
It has been discovered by Schay \cite{schay} and Dudley
\cite{dudley} that Kramers phase space diffusion admits
relativistic generalization. The generalization is unique under
the assumption that the generator depends only on the phase space
variables $(x,p)$ and that the particle mass does not change
during the random evolution. It seems that the authors of
ref.\cite{malik} unaware of the work of Schay and Dudley have
introduced another definition of the relativistic diffusion which
depends on an additional four-vector interpreted as a fluid
velocity (without this four-vector the model would not be
covariant). This approach is developed in \cite{deb}\cite{deb2}.
In another approach \cite{hang1}-\cite{hang2}  the authors
following methods previously applied to  the non-relativistic
Langevin equation obtain the models of
refs.\cite{malik}\cite{deb2} as well as the model of Schay and
Dudley. We have discussed
  \cite{haba} drifts added to  the relativistic diffusion of Schay and Dudley in such a way
 that an equilibrium is achieved and the detailed balance
 satisfied. This model has been independently
 discovered and studied in detail in \cite{calo}.
 The existence of an equilibrium which is the J\"uttner
 distribution \cite{juttner} has been also shown in the models of refs.\cite{malik}\cite{deb2}
 \cite{hang1}. These diverse investigations did not show
 that there is a physical model realizing a particular version of
 the relativistic diffusion. The   model of
 Schay and Dudley is appealing by its mathematical elegance
 (it has a generalization to arbitrary pseudoriemannian manifolds
 \cite{lejan}). Nevertheless,  it has been rejected in \cite{deb}\cite{deb2}
 as unphysical. In this paper we  show that the
 relativistic diffusion of Schay and Dudley results from a
 dynamical model of a particle motion in a random electromagnetic field
 in the same way as the non-relativistic Brownian motion
 \cite{chandra} can be derived from a non-relativistic dynamics in
 the electromagnetic field. Moreover, we show that a particular version of the
 relativistic Ornstein-Uhlenbeck process suggested in \cite{haba}
and  \cite{calo} (see also \cite{hang2}) comes  from an interaction
with the relativistic random electromagnetic field.

In non-relativistic mechanics the diffusive dynamics can be
derived from the dynamics in a random force
\cite{kesten}\cite{komor}\cite{zheng1}\cite{zheng2} (these papers
describe also the history of the problem with proper citations).
The random force felt by a tracer particle can be understood as
the force coming from a chaotic motion of other particles. The
motion in a random electromagnetic field has been studied by
physicists for a long time because of its relevance to
astrophysics,plasma physics and high-energy physics \cite{chandra}
\cite{jok}\cite{stu1}\cite{stu2}\cite{ich}\cite{svetitsky}. The
notion of a random Liouville operator has been introduced by Kubo
\cite{kubo1}. In his paper it has been shown that the Markov
approximation to the random motion leads to the diffusion (for a
rigorous proof of the non-relativistic diffusion see
\cite{kesten}\cite{komor}). In view of the results obtained for
non-relativistic dynamics it is suggestive to repeat the
calculations in the relativistic case in order to see the
appearance of the diffusive behaviour in a relativistic motion.

 The plan of this paper is as follows. In
sec.2 we define the relativistic  dynamics of a massive particle
in an electromagnetic field in the proper time and in the
laboratory (coordinate) time. We move to a statistical description
of classical dynamics in terms of the evolution of functions on
the phase space  generated by the Liouville first order
differential operator. Such a formalism is a preparation for a
unifying description of deterministic and random systems.
 In sec.3 we define a random Gaussian antisymmetric tensor field
 $F_{\mu\nu}$ which can be considered as an electromagnetic field whose
 randomness comes from random sources. To the best of author's
 knowledge the relativistic random electromagnetic field is
 constructed here for the first time.
 It could be considered as a regular(soft) version of quantum
electromagnetic field when the short distance singularities of QED
are ignored.  We discuss an evolution of observables generated by
(an adjoint of ) a random Lorentz invariant Liouville operator
(sec.4) in the sense of Kubo \cite{kubo1}. We represent the
evolution in a form suitable for averaging over the electromagnetic
field. We refer to Kubo \cite{kubo1}-\cite{kubo2} for an argument
that it is sufficient to calculate the expectation value of the
square of the Liouville operator in order to determine the generator
of the diffusion which results from the Markov approximation of the
random dynamics. The calculation of an expectation value of the
square of the Liouvile operator generating the proper time dynamics
is performed in sec.5. It comes out that till the second order in
the electromagnetic field the random dynamics coincides with the
diffusive dynamics resulting from the diffusion generator of Schay
and Dudley. This is the relativistic analog of the results derived
for the non-relativistic motion in
refs.\cite{jok}\cite{stu1}\cite{stu2}. It shows that for a particle
motion in a random field the relativistic diffusion plays the same
role as the classical Brownian motion.

 We perform the same calculations in the
laboratory time (sec.6). Now, we do not have the explicit Lorentz
invariance of the Liouville operator. An average of the square of
the Liouville operator over the electromagnetic field gives a
surprise. There appears a first order Lorentz non-invariant term.
We introduced such a  drift  in \cite{haba} as the friction
determined by the J\"uttner equilibrium distribution
\cite{juttner} through the detailed balance condition. The
J\"uttner distribution which comes from the random dynamics
corresponds to the inverse temperature $\beta^{-1}=mc^{2}$ ($m$ is
the particle's mass). In the controversy concerning the
relativistic analog of the Ornstein-Uhlenbeck process as discussed
in \cite{deb}\cite{hang}\cite{malik} \cite{deb2}the interaction
with a random electromagnetic field selects the process determined
by the detailed balance from the diffusion of Schay and Dudley.

\section{Relativistic dynamics}
  The dynamics of a particle with mass $m$
in an external electromagnetic field neglecting its own
electromagnetic field (produced by the current of a moving
particle)  is described by the explicitly Lorentz covariant
four-vector equations \cite{landau}
 \begin{equation}
 \frac{dx^{\mu}}{d\tau}=\frac{1}{mc}p^{\mu},
 \end{equation}
 \begin{equation}
 mc\frac{dp_{\mu}}{d\tau}=F_{\mu\nu}p^{\nu},
 \end{equation}
The vectors are defined in the Minkowski space-time with the
metric tensor $\eta^{\mu\nu}=(1,-1,-1,-1)$, the Greek index
$\mu=0,1,2,3$. We shall also use Latin indices $j=1,2,3$ . It
follows from eq.(2) that
\begin{equation}
\frac{d}{d\tau}(\eta^{\mu\nu}p_{\mu}p_{\nu}) =0. \end{equation}
Hence,\begin{equation}
p^{2}=\eta^{\mu\nu}p_{\mu}p_{\nu}=m^{2}c^{2},
\end{equation}From eqs.(1) and (4)
it follows that $\tau$ is the proper time
\begin{equation} d\tau^{2}=dx^{\mu}dx_{\mu}.
\end{equation}
We can eliminate $\tau$ from eqs.(1)-(2) in favor of $x^{0}$ (we
call $x^{0}$ the laboratory time). Then, eqs.(1)-(2) read
\begin{equation}
 \frac{dx^{k}}{dx^{0}}=\frac{1}{p_{0}}p^{k},
 \end{equation}
 \begin{equation}
 \frac{dp_{k}}{dx^{0}}=F_{k\nu}p^{\nu}p_{0}^{-1}
 \end{equation}$k=1,2,3.$

We move from a study of individual trajectories to functions $ W$
(observables) on the phase space $(x,p)$. Then, $W$ evolves as
\begin{equation}\partial_{\tau}W=\frac{p^{\mu}}{mc}\frac{\partial W}{\partial
x^{\mu}}-F_{j\nu}\frac{p^{\nu}}{mc}\frac{\partial W}{\partial
p^{j}}.\end{equation} There is no derivative over $p^{0}$ in
eq.(8) as $p^{0}$ is expressed by ${\bf p}$. Knowing the evolution
of $W$ we can obtain the trajectory $(x(\cdot),p(\cdot))$. It is
useful to apply a statistical formulation of classical dynamics.
We begin with a probability density $\Omega$ of the initial points
$(x,p)$ in the phase space and determine its evolution in time
$\tau$ . We define an expectation value of the observable $W$ in a
state (probability distribution) $\Omega $ as
\begin{equation}
\Omega(W)=(\Omega,W)=\int dxd{\bf p}\Omega W
\end{equation}
Then, the adjoint evolution is
\begin{equation}\begin{array}{l}
(\Omega_{\tau},W)\equiv(\Omega,W_{\tau}).\end{array}
\end{equation}
The initial probability distribution $\Omega$ in the phase space
becomes a probability distribution of trajectories
$(x(\cdot),p(\cdot))$ satisfying the equation
\begin{equation}\begin{array}{l}\partial_{\tau}\Omega=-\frac{p^{\mu}}{m}\frac{\partial\Omega}{\partial
x^{\mu}}+F_{j\nu}\frac{p^{\nu}}{mc}\frac{\partial\Omega}{\partial
p^{j}}.
\end{array}\end{equation} The evolution of an observable $\psi$ in
the laboratory time is determined by eqs. (6)-(7)
\begin{equation}\frac{\partial\psi}{\partial
x^{0}}={\bf p}p_{0}^{-1}\nabla_{{\bf x}} \psi
+F_{j\nu}p^{\nu}p_{0}^{-1}\frac{\partial\psi}{\partial
p^{j}}.\end{equation} By means of eq.(10) we can again define an
evolution of the probability distribution. Such a statistical
description is necessary when we consider the random dynamics
because in such a case an individual trajectory is not an
observable. Only averages over trajectories have a physical
meaning.
 \section{Random electromagnetic fields}

 We consider
an antisymmetric tensor field $F$ which is covariant with respect
to the Poincare group. We assume that an average  $\langle\cdot
\rangle$ over $F$ is defined which preserves the Poincare
symmetry. This means that the two-point function defined by
\begin{equation}\langle
F_{\mu\nu}(x)F_{\sigma\rho}(x^{\prime})\rangle=G_{\mu\nu
;\sigma\rho}(x-x^{\prime}).
\end{equation}
is a tensor depending only on $x-x^{\prime}$ with the following
symmetry properties:

i)$G_{\mu\nu ;\sigma\rho}$ is symmetric under the exchange of
indices $(\mu\nu;x)$ and $(\sigma\rho;x^{\prime})$ and
antisymmetric under the exchange $\mu\rightarrow\nu$ and
$\sigma\rightarrow\rho$

We impose the first set of Maxwell equations (Bianchi identities)
on the tensor field $F$
\begin{equation}
\partial_{\mu} \epsilon^{\mu\nu\sigma\rho}F_{\sigma\rho}=0.\end{equation}
 which imply an equation for the covariance (13)

ii)
\begin{equation}
\partial_{\alpha}\epsilon^{\alpha\beta\mu\nu}G_{\mu\nu
;\sigma\rho}=0.
\end{equation}
In Fourier transforms ($\tilde{F}(-k)=\overline{\tilde{F}}(k)$
because $F$ is real) eq.(13) reads
\begin{equation}\langle\overline{\tilde{F}}_{\mu\nu}(k)\tilde{F}_{\sigma\rho}(k^{\prime})\rangle=
\tilde{G}_{\mu\nu ;\sigma\rho}(k)\delta(k-k^{\prime}),
\end{equation}
where $\tilde{G}_{\mu\nu ;\sigma\rho}(k)$ is a tensor which must
be constructed from the vector $k_{\mu}$ and the fundamental
four-dimensional tensors $\eta_{\mu\rho}$ and
$\epsilon_{\mu\nu\rho\sigma}$. Hence, the tensor
$\tilde{G}_{\mu\nu ;\sigma\rho}$ with the symmetries i) has the
form
\begin{equation}\begin{array}{l}
\tilde{G} _{\mu\nu
;\sigma\rho}(k)=a_{1}(k)(\eta_{\mu\sigma}k_{\nu}k_{\rho}
-\eta_{\mu\rho}k_{\nu}k_{\sigma}+ \eta_{\nu\rho}k_{\mu}k_{\sigma}
-\eta_{\nu\sigma}k_{\mu}k_{\rho})\cr+a_{2}(k)(\eta_{\mu\sigma}\eta_{\nu\rho}
-\eta_{\mu\rho}\eta_{\nu\sigma})+a_{3}(k)\epsilon_{\mu\nu\sigma\rho},
\end{array}\end{equation} where $a_{j}$ are scalars with respect to the Lorentz group.  Eq.(15) after the Fourier transform reads
\begin{equation}
k_{\alpha}\epsilon^{\alpha\beta\mu\nu}\tilde{G} _{\mu\nu
;\sigma\rho}(k)=0.
\end{equation}
It follows from eqs.(17) and (18) that $a_{1}$ can be an arbitrary
 function of $k^{2}$ and $a_{2}=0$. From eq.(18) we
obtain an equation for $a_{3}$
\begin{equation}
k_{\alpha}\epsilon^{\alpha\beta\mu\nu}\epsilon_{\mu\nu\sigma\rho}a_{3}(k)=0.
\end{equation}
Hence, $a_{3}(k)=0 $ if $k\neq 0$ (i.e. $a_{3}=K\delta(k)$). In
the configuration space the admissible $a_{3}$ term would lead to
a
 constant term ($K$ is a constant)
 \begin{equation}G^{\epsilon}_{\mu\nu ;\sigma\rho}=K
 \epsilon_{\mu\nu\sigma\rho}.\end{equation}

So far we did not make any assumption concerning the question
where the average value (13) comes from. As an example, it could
come from a time average. It may result from a certain probability
distribution of the currents
\begin{equation}
\partial^{\mu}F_{\mu\nu}=J_{\nu}
\end{equation} or the electromagnetic field
becomes random as a result of a passage through a random medium.
In the latter cases
  there exists a probability measure
 determining the distribution of $F_{\mu\nu}$. It is known
 \cite{gelfand} that a necessary  condition for an existence of
 such a measure  is

\begin{equation}
\langle \Big(\int dx F_{\mu\nu}(x)f^{\mu\nu}(x)\Big)^{2}\rangle
=\int
dxdx^{\prime}f^{\mu\nu}(x)f^{\sigma\rho}(x^{\prime})G_{\mu\nu;\sigma\rho}(x-x^{\prime})\geq
0
\end{equation}
for arbitrary test-functions $f^{\mu\nu}$ (satisfying some
regularity and growth conditions \cite{gelfand}). The condition (22)
is also sufficient for an existence of a random Gaussian field $F$.

The positivity condition (22) excludes the $\epsilon$-term (20)
because the quadratic form
\begin{displaymath}
\int dx
dx^{\prime}f^{\mu\nu}(x)f^{\sigma\rho}(x^{\prime})\epsilon_{\mu
\nu\sigma\rho} \end{displaymath}cannot be positive for arbitrary
test-functions $f^{\mu\nu}$. It follows that only the $a_{1}$
term, which is a function of $k^{2}$, remains in eq.(17). Hence,
\begin{equation}
\langle F_{\mu\nu}(x)F_{\sigma\rho}(x^{\prime})\rangle=-D_{\mu\nu
;\sigma\rho}G(x-x^{\prime}),\end{equation}
 where\begin{equation}\begin{array}{l}
 D_{\mu\nu;\sigma\rho}
=-\eta_{\mu\sigma}\partial_{\nu}\partial_{\rho}+
\eta_{\mu\rho}\partial_{\nu}\partial_{\sigma}-
\eta_{\nu\rho}\partial_{\sigma}\partial_{\mu}+
\eta_{\nu\sigma}\partial_{\mu}\partial_{\rho}
\end{array}\end{equation}and $G$ is a  function depending only on $(x-x^{\prime})^{2}$.

There remains to explore the positivity condition (22). We show
that under the assumptions i)-ii)
   there exists a random  antisymmetric tensor field with the
  two-point correlation function (23)  if and only if the Fourier transform $\tilde{G}$ of the function
  $G$ in eq.(23) satisfies the condition

iii) \begin{equation} \tilde{G}(k)\geq 0
\end{equation} and $\tilde{G}(k)$=0 if $k^{2}< 0$.

We prove that the inequality (22) follows from the condition iii).
With the two-point correlation function (23) eq.(22) reads
\begin{equation}\begin{array}{l} \int
dxdx^{\prime}f^{\mu\nu}(x)f^{\sigma\rho}(x^{\prime})G_{\mu\nu;\sigma\rho}(x-x^{\prime})
\cr=-\int dk \tilde{f}^{\mu\nu}(-k)(\eta_{\mu\sigma}k_{\nu}k_{\rho}
-\eta_{\mu\rho}k_{\nu}k_{\sigma}+ \eta_{\nu\rho}k_{\mu}k_{\sigma}
-\eta_{\nu\sigma}k_{\mu}k_{\rho})\tilde{f}^{\sigma\rho}\tilde{G}(k)
\cr=\int dk \overline{g^{j}(k)}g^{j}(k)\tilde{G}(k)-\int
dk\overline{k_{j}\tilde{f}^{0j}}k_{l}\tilde{f}^{0l}(k)\tilde{G}(k),\end{array}\end{equation}
where \begin{equation} g^{j}
=k_{0}\tilde{f}^{0j}+k_{l}\tilde{f}^{lj} .\end{equation} Expressing
$k_{j}\tilde{f}^{0j}$ by $k_{j}g^{j}$ we obtain
\begin{equation}\begin{array}{l}
\int
dxdx^{\prime}f^{\mu\nu}(x)f^{\sigma\rho}(x^{\prime})G_{\mu\nu;\sigma\rho}(x-x^{\prime})
=\int dk \tilde{G}(k)\overline{ g^{j}}g^{j}-\int
dk\tilde{G}(k)k_{0}^{-2}\vert g^{j}k_{j}\vert^{2}\cr\geq \int
dk\tilde{G}(k)k_{0}^{-2}\overline{g^{j}}g^{j}(k_{0}^{2}-{\bf
k}^{2})\geq 0,
\end{array}\end{equation} if (25) and the condition iii) are satisfied.

{\bf Remarks}:

1.Note that there is a function $M_{\nu\sigma\rho}$ such
 that
\begin{displaymath}
\langle
\partial^{\mu}F_{\mu\nu}(x)F_{\sigma\rho}(x^{\prime})\rangle=\partial^{\mu}\partial_{\mu}M_{\nu\sigma\rho}(x-x^{\prime}).
\end{displaymath}
We treat this consequence of eq.(23) as a weak form of
 the second set
of Maxwell equations. The condition that the electromagnetic field
has no source ($J=0$) can now be imposed by the requirement that
$M$ satisfies the wave equation .

2. If the source-less Maxwell equations (21) are satisfied then
\begin{equation}
\tilde{G}(k)=\delta(k^{2})
 \end{equation}
(eq.(29), with $k_{0} \geq 0$, is satisfied for the vacuum
two-point function in quantum field theory of the free
electromagnetic field \cite{schweber}). However, the two-point
function (23) determined by its Fourier transform (29) is
singular. The singularity would appear in the diffusion equation
as a singularity of the diffusion coefficients. We do not impose
$J=0$ in eq.(21) in order to work with more regular
electromagnetic fields.

 It follows from eqs.(23)-(24) that
\begin{equation}\begin{array}{l} D_{\mu\nu;\sigma\rho}G
=\eta_{\mu\sigma}G_{\nu\rho}- \eta_{\mu\rho}G_{\nu\sigma}+
\eta_{\nu\rho}G_{\sigma\mu}- \eta_{\nu\sigma}G_{\mu\rho}
\end{array}\end{equation}
with
\begin{equation}
G_{\mu\nu}(x)\equiv\partial_{\mu}\partial_{\nu}G\equiv\eta_{\mu\nu}h_{1}(x)+x_{\mu}x_{\nu}h_{2}(x).
\end{equation}
Here, $G(x)\equiv g(u)$ with $u=x^{2}$, $h_{1}(x)=2g^{\prime}(u)$
and $h_{2}(x)=4g^{\prime\prime}(u)$. The functions $h_{j}(x)$
depend only on $x^{2}=x_{\mu}x^{\mu}$.

If in a model of an electromagnetic field the scalar $G$ is a
function of some other vectors (in addition to $x$) then G is a
function of  scalars built  from these vectors. In such a case the
 formula (31) contains more terms. As a consequence, the diffusion
discussed in
 subsequent sections would also depend on the additional
 vectors.

\section{Random evolution } In this section we discuss a solution
of the evolution equations (8) and (11). We are going to represent
the solution in such a form that an averaging over the
electromagnetic field can be performed. The differential equations
(8) and (11) for the evolution of functions $W$ on the phase space
are of the form
\begin{equation}
\partial_{s}W_{s}=(X+Y)W_{s},
\end{equation}
where $Y$ is a random first order differential operator and $X$ is
the free evolution. For the evolution (8) in the proper time
\begin{equation} X=\frac{1}{mc}p^{\mu}\partial_{\mu}.
\end{equation}
Let
\begin{equation}
Y(s)=\exp(-sX)Y\exp(sX) .\end{equation} Then, the solution of
eq.(32) can be expressed as
\begin{equation}
W_{t}=\exp(tX)W^{I}_{t},
\end{equation}
where
\begin{equation}
\partial_{s}W^{I}_{s}=Y(s)W^{I}_{s},
\end{equation}(this is the interaction picture well-known from
quantum mechanics). We can solve eq.(36) by iteration. The
iteration till the second order reads
\begin{equation}
\begin{array}{l}
W_{t}^{I}=W_{0}+\int_{0}^{t}dsY(s)W_{s}^{I}
+\int_{0}^{t}ds\int_{0}^{s}ds^{\prime}Y(s)Y(s^{\prime})W_{s^{\prime}}^{I}\cr
=W_{0}+\int_{0}^{t}dsY(s)W_{s}^{I}
+\frac{1}{2}\int_{0}^{t}ds\int_{0}^{s}ds^{\prime}(Y(s)Y(s^{\prime})+Y(s^{\prime})Y(s)+
[Y(s),Y(s^{\prime})])W_{s^{\prime}}^{I}.
\end{array}\end{equation}We can expand the solution (37) in an
infinite series. We assume that the initial condition $W$ for
eq.(32) is independent of the electromagnetic field. Then, the
expectation values of the products of $Y(s)$ appearing in the series
can be calculated because the correlation functions of the Gaussian
random field $F$ are determined by the two-point function (23).
 In the form of
a path-ordered integral the solution has the form
\begin{equation}
W_{t}^{I}=T\Big(\exp\Big(\int_{0}^{t}dsY(s)\Big)\Big)W_{0}.
\end{equation}
The diffusion limit applied in secs.5-6 relies on the
approximation \cite{kubo1}\cite{kubo2}
\begin{equation}
\langle
T\Big(\exp(\int_{0}^{t}dsY(s))\Big)\rangle=\exp\Big(\frac{1}{2}\int_{0}^{t}ds\int_{0}^{s}ds^{\prime}
\langle (Y(s)Y(s^{\prime})+Y(s^{\prime})Y(s))\rangle\Big).
\end{equation}
If $[Y(s),Y(s^{\prime})]=0$ and $Y$ is a linear function of
Gaussian variables then eq.(39) is exact. In our case the operator
 $Y(s)$ (generating the Liouville dynamics) is an antihermition
operator in $L^{2}(\frac{d{\bf p}}{p_{0}}dx)$. Then, the
anticommutator $Y(s)Y(s^{\prime})+Y(s^{\prime})Y(s)$ is a hermition
operator whereas  $[Y(s),Y(s^{\prime})]$ is antihermition. We
neglect the commutator in eqs.(37) and (39) corresponding to the
unitary Liouville evolution and restrict ourselves to the pure
diffusion.
\section{The expectation value of the proper time evolution} The
observable time evolution results from an averaging over
fluctuations of the electromagnetic field. The fluctuations of the
electromagnetic field make the dynamics chaotic. The particles'
trajectories become random.  In this section we apply the approach
of Kubo \cite{kubo1}\cite{kubo2} in order to obtain a diffusion as
a Markov approximation to random dynamics. The result involves an
approximation of the Liouville evolution of functions on the phase
space by an evolution generated by  second order differential
operators. In this formulation it is sufficient to calculate an
expectation value of the square of the Liouville operator in order
to determine the generator of the diffusion.
 We consider the proper
time evolution (1)-(2) first. In general, we could split the
Liouville operator $Y^{tot}=Y^{ex}+Y$, where $Y^{ex}$ is the
Liouville operator corresponding to an external deterministic
electromagnetic field and $Y$ describes the random part. The
$Y^{ex}$ part could be added to the final result. We restrict our
discussion to the random Liouville operator (34). Taking as $Y$
the second term in eq.(8) and $X$ from eq.(33) we calculate $Y(s)$
(34)
\begin{equation}
Y(s)=F_{j\nu}(x-\frac{s}{mc}p)p^{\nu}\frac{\partial}{\partial
p^{j}}.\end{equation} We apply the covariance (30)-(31) (with
$x^{\mu}\simeq \frac{s}{mc}p^{\mu}$ from eq.(40)) to calculate the
expectation value  appearing in eq.(39)
\begin{equation}\begin{array}{l}
\frac{1}{2}\int_{0}^{\tau}ds\int_{0}^{s}ds^{\prime}\langle
(Y(s)Y(s^{\prime})+Y(s^{\prime})Y(s))\rangle=
(mc)^{-2}\int_{0}^{\tau}ds\int_{0}^{s}ds^{\prime} \cr
\Big(\eta_{jl}(\eta_{\nu\rho}H_{1}(s-s^{\prime})+m^{-2}c^{-2}p_{\nu}p_{\rho}H(s-s^{\prime}))
\cr -\eta_{j\rho}(\eta_{\nu
l}H_{1}(s-s^{\prime})+m^{-2}c^{-2}p_{\nu}p_{l}H(s-s^{\prime})) \cr +
\eta_{\nu\rho}(\eta_{jl}H_{1}(s-s^{\prime})+m^{-2}c^{-2}p_{j}p_{l}H(s-s^{\prime}))
\cr -\eta_{\nu
l}(\eta_{j\rho}H_{1}(s-s^{\prime})+m^{-2}c^{-2}p_{j}p_{\rho}H(s-s^{\prime}))\Big)
p^{\nu}\frac{\partial}{\partial p^{j}}
p^{\rho}\frac{\partial}{\partial p^{l}},
\end{array}\end{equation}
where \begin{displaymath}H_{1}(s-s^{\prime})=
h_{1}((s-s^{\prime})^{2})=2g^{\prime}(u)\end{displaymath} and
\begin{displaymath}H(s-s^{\prime})=(s-s^{\prime})^{2}h_{2}((s-s^{\prime})^{2})
=4ug^{\prime\prime}(u).\end{displaymath}where
$u=(s-s^{\prime})^{2}$, $h_{j}(x)$ (defined after eq.(31)in terms
of $G(x^{2})\equiv g(\sqrt{x^{2}})$) being functions of $x^{2}$
depend only on $s-s^{\prime}$ because after a calculation of the
two-point function (13) of $F(x-\frac{s}{mc}p)$ and
$F(x-\frac{s^{\prime}}{mc}p)$ we obtain
\begin{displaymath}
h_{j}\Big(((s-s^{\prime})p(mc)^{-1})^{2})=h_{j}((s-s^{\prime})^{2}).
\end{displaymath}
The explicit calculations in eq.(41) give
\begin{equation}\begin{array}{l}
\frac{1}{2}\int_{0}^{\tau}ds\int_{0}^{s}ds^{\prime}\langle
(Y(s)Y(s^{\prime})+Y(s^{\prime})Y(s))\rangle =
(mc)^{-2}\int_{0}^{\tau}ds\int_{0}^{s}ds^{\prime}
\Big(2H_{1}(s-s^{\prime})+H(s-s^{\prime})\Big )\triangle_{H}^{m},
\end{array}\end{equation}
where \begin{equation}\begin{array}{l} \triangle_{H}^{m}
=(\delta^{jl}m^{2}c^{2}+p^{j}p^{l})\frac{\partial}{\partial
p^{l}}\frac{\partial}{\partial
p^{j}}+3p^{l}\frac{\partial}{\partial p^{l}}\end{array}
\end{equation}
 is the generator of the relativistic diffusion of
Schay \cite{schay} and Dudley \cite{dudley}. In the approximation
(39) (neglecting the non-commutativity) the result of an average
over a random electromagnetic field is the relativistic diffusion
with the time transformed by the function standing in front of
$\triangle_{H}^{m}$ in eq.(42). Note that $\triangle_{H}^{m}$ is
self-adjoint in $L^{2}(d{\bf p}p_{0}^{-1})$ similarly as its
non-relativistic analog (generator of Krammers diffusion) is
self-adjoint in $L^{2}(d{\bf p})$.
\section{Random evolution in laboratory time}
Eqs.(1)-(2) are equivalent to eqs.(6)-(7). Both sides of
eqs.(1)-(2) are four-vectors transforming linearly under the
Lorentz group. This property does not hold true for eqs.(6)-(7).
The equivalence between the description in terms of the proper
time and coordinate time may be lost if we perform non-linear
transformations such as the averaging over random fields. For this
reason we calculate the expectation values of the square of the
laboratory time Liouville generator (12) again. In the laboratory
time the free evolution is determined by
\begin{equation} X=p_{0}^{-1}{\bf p}\nabla_{{\bf x}}.
\end{equation}
Then from eqs.(12) and (34)
\begin{equation}
Y(s)=F_{l\mu}({\bf x}-sp_{0}^{-1}{\bf
p},s)p^{\mu}p_{0}^{-1}\frac{\partial}{\partial
p^{l}}.\end{equation}  For the calculation of  the expectation
values of $Y$ (45) we must make the replacements in formulas of
sec.5
\begin{displaymath}
p\rightarrow p_{0}^{-1}{\bf p},\end{displaymath}
\begin{displaymath}
{\bf x}\rightarrow p_{0}^{-1}{\bf p}s\end{displaymath} and
$x_{0}-x^{\prime}_{0}=s-s^{\prime}$. So that \begin{equation}
(x-x^{\prime})^{2}=p_{0}^{-2}m^{2}c^{2}(s-s^{\prime})^{2}
\end{equation}
as the argument of the functions $h_{j}$ in eq.(41). From eq.(46)
we can see that the transition from the proper time evolution to
the laboratory time evolution involves ${\bf p}\rightarrow {\bf
p}p_{0}^{-1}$. This is equivalent to $s\rightarrow smcp_{0}^{-1}$
in the functions $h_{j}$ when we calculate the expectation value
(41). Changing the time integration variables $s\rightarrow
smcp_{0}^{-1}$ in eq.(41) we obtain (we write $x_{0}=ct$)
\begin{equation}\begin{array}{l}
\frac{1}{2}\int_{0}^{ct}ds\int_{0}^{s}ds^{\prime}\langle
(Y(s)Y(s^{\prime})+Y(s^{\prime})Y(s))\rangle
\cr=\int_{0}^{tmc^{2}p_{0}^{-1}}ds\int_{0}^{s}ds^{\prime}
\Big(\eta_{jl}(\eta_{\nu\rho}H_{1}(s-s^{\prime})+m^{-2}c^{-2}p_{\nu}p_{\rho}H(s-s^{\prime}))
\cr -\eta_{j\rho}(\eta_{\nu
l}H_{1}((s-s^{\prime})+m^{-2}c^{-2}p_{\nu}p_{l}H(s-s^{\prime}))\cr +
\eta_{\nu\rho}(\eta_{jl}H_{1}((s-s^{\prime})+m^{-2}c^{-2}p_{j}p_{l}H(s-s^{\prime}))
\cr -\eta_{\nu
l}(\eta_{j\rho}H_{1}(s-s^{\prime})+m^{-2}c^{-2}p_{j}p_{\rho}H(s-s^{\prime}))\Big)
p_{0} p^{\nu}\frac{\partial}{\partial p^{j}}
p^{\rho}p_{0}^{-1}\frac{\partial}{\partial p^{l}}.
\end{array}\end{equation}
We have an additional term $D_{1}$ (coming from $p^{j}$
differentiation of $p_{0} ^{-1}$ on the rhs of eq.(47)) in
comparison to the rhs of eq.(41)
\begin{equation}D_{1}=-p_{0}^{-1} p^{\nu}p^{j}\frac{\partial}{\partial
p^{l}}.
\end{equation}
This term after a contraction with the  tensors in eq.(47) gives
the result
\begin{equation}\begin{array}{l}
\frac{1}{2}\int_{0}^{ct}ds\int_{0}^{s}ds^{\prime}\langle
(Y(s)Y(s^{\prime})+Y(s^{\prime})Y(s))\rangle
=\int_{0}^{tmc^{2}p_{0}^{-1}}ds\int_{0}^{s}ds^{\prime}
\Big(2H_{1}(s-s^{\prime})+H(s-s^{\prime})\Big) \triangle_{\beta}
\end{array}\end{equation}
with\begin{equation} \triangle_{\beta}=\triangle^{m}_{H}-p_{0}
p^{j}\frac{\partial}{\partial p^{j}},
\end{equation}
where the last term in eq.(50) is the friction introduced in
\cite{haba}. $\triangle_{\beta}$ generates a diffusion which
equilibrates to the J\"uttner distribution with the inverse
temperature $\beta^{-1}=mc^{2}$. We obtain the J\"uttner
equilibrium distribution \cite{juttner} $\Phi_{E}=\exp(-\beta
cp_{0})$ from the requirement
\begin{displaymath}
\triangle_{\beta}^{*}\Phi_{E}=0,
\end{displaymath}
where the adjoint is in $L^{2}(d{\bf p})$ (see eq.(9)).

 It is surprising that the proper time dynamics (1)-(2) and the
 laboratory time dynamics (6)-(7) after an average over the
 electromagnetic field give inequivalent results.
The probability distribution $\Omega$ of the random dynamics in
the laboratory time tends to the
 J\"uttner  equilibrium distribution $\Phi_{E}$ when time goes to infinity. The covariant dynamics of
 sec.5 has no limit when the proper time or the laboratory time go to infinity.
It is easy to comprehend the difference from the technical point
of view as the Liouville generator in  laboratory time dynamics
has the $p_{0}^{-1}$ factor which leads to the extra term (48).
However, some mysteries remain concerning the question why
equivalent dynamics have different diffusive limits.

 We assume  that  the functions $H_{1}$ and $H$ decay fast for $s\neq
 s^{\prime}$. We
approximate the $s^{\prime}$-integral  appearing in eqs.(42) and
(49) as follows\begin{equation}\begin{array}{l}
\int_{0}^{s}ds^{\prime}
\big(2H_{1}(s-s^{\prime})+H(s-s^{\prime}))\simeq\int_{0}^{\infty}ds^{\prime}
\big(2H_{1}(s^{\prime})+H(s^{\prime}))=\frac{\kappa^{2}}{2},
\end{array}\end{equation} where
$\kappa^{2}$ is the diffusion constant of ref.\cite{haba}. We
apply the formulae below eq.(41), expressing $H_{1}$ and $H$ in
terms of $G(x)$ to calculate

\begin{equation}
\kappa^{2}=2\int_{0}^{\infty}duu^{-\frac{1}{2}}g^{\prime}(u)
\end{equation}
,where $G(x)\equiv g(u)$ with $u=x^{2}$.

The  Markov approximation must be performed also for higher order
terms in the expansion (37) in order to  justify the formula (39).
The approximation has been discussed first by Kubo
\cite{kubo1}-\cite{kubo2} (the expression of the diffusion constant
(51) and (52) by the correlation function of the forces $F_{\mu\nu}$
defining the functions $h_{j}$  is known as the Kubo formula).
 Summarizing
the results (42) and (49) we  have in the Markov approximation
\begin{equation}
\partial_{\tau}W_{\tau}^{I}=\frac{\kappa^{2}}{2}\triangle_{H}^{m}W_{\tau}^{I}
\end{equation}
When we define $W$ as  in eq.(35) then  $W$ satisfies the equation
\begin{equation}
\partial_{\tau}W_{\tau}=p^{\mu}\frac{\partial}{\partial
x^{\mu}}W_{\tau}+\frac{\kappa^{2}}{2}\triangle_{H}^{m}W_{\tau}.
\end{equation}
In the laboratory time the corresponding equations read
\begin{equation} p_{0}\frac{\partial}{\partial
x_{0}}W_{x_{0}}^{I}=\frac{\kappa^{2}}{2}\triangle_{\beta}W_{x_{0}}^{I}.
\end{equation}and
\begin{equation}
p^{\mu}\frac{\partial}{\partial
x^{\mu}}W=\frac{\kappa^{2}}{2}\triangle_{\beta}W.
\end{equation}

\section{Summary and outlook}
We have been studying a model of particle relativistic dynamics in
a random electromagnetic field showing diffusive behaviour. The
aim was to find which of the models of the relativistic diffusion
encountered in the literature is realized in this physical model.
For this reason we have calculated an expectation value of the
square of the Liouville operator which defines the generator of
the diffusion. In the case of the dynamics in proper time, as the
result of the calculations, we obtain the generator  of the
diffusion of Schay and Dudley; a mathematical model discovered a
long time ago but disregarded by most physicists. We repeat the
calculations for the dynamics in the laboratory time. As a result
we obtain the generator of the proper time diffusion plus a drift
term which moves the probability distribution to the equilibrium
(the J\"uttner equilibrium distribution at $\beta^{-1}=mc^{2}$).
Comparing with the non-relativistic dynamics in the
electromagnetic field we can see that the diffusive dynamics in
the proper time can be treated as a relativistic analog of the
Brownian motion whereas the dynamics in the laboratory time is
analogous to the one of the Ornstein-Uhlenbeck process. It follows
that the diffusion of Schay and  Dudley should be treated as a
starting point for  an investigation of the  diffusive motion of
relativistic systems in the same way as the Brownian motion (in
Krammers version) serves for a construction of other
non-relativistic diffusions in physics \cite{chandra} as well as
in mathematics.

Some extensions of our formulation can be suggested. Applying the
Martin-Siggia-Rose formalism
\cite{martin}\cite{zheng1}\cite{zheng2} one could obtain a
functional measure (defined on trajectories) corresponding to the
motion in a random electromagnetic field. The exact formula for
the functional measure can be applied for further investigations
of the random motion beyond the Markov approximation. It could
also be used for an inquiry of the meaning of the proper time
dynamics in random relativistic systems and the relation to
laboratory time dynamics. The equilibration to the J\"uttner
distribution (expected on physical grounds) resulting from the
generator (50) may indicate a deeper mathematical mechanism (in
the path integral formalism) leading to the relativistic version
of the Ornstein-Uhlenbeck process. In another direction, we can
assume a dependence of the two-point function (13) on some other
tensors (e.g. a vector describing a moving frame for theories at
finite temperature). In such a case the formalism does not change.
We can calculate the dependence of the resulting diffusion (which
can be treated as a perturbation of the one studied here) on the
additional variables. Then, astrophysical applications of the
relativistic diffusion require a general pseudoriemannian metric
(as in \cite{lejan}). The formulation of secs.2-6 can  be extended
to general relativity. A local Markov approximation leading to the
diffusion on a manifold is still possible. However, a global
approximation of the relativistic dynamics  on a manifold by the
diffusion of ref.\cite{lejan} may encounter some difficulties.

\end{document}